\documentstyle[psfig]{l-aa}
\begin{document}

\thesaurus{04.01.2, 08.01.2, 08.12.1, 08.06.3, 08.07.1}

\title {Library of high-resolution UES echelle spectra 
 of F, G, K and M field dwarf stars
\thanks{Based on observations made with the William Herschel Telescope
operated on the island of La Palma by the Isaac Newton Group at the 
Spanish Observatorio del Roque de
Los Muchachos of the Instituto de Astrof\'{\i}sica de Canarias}
\thanks{
Fig. 1 and Tables 1 to 5 are only available in electronic form, 
and Table 6 is also available in electronic form 
}             
\thanks{The spectra of the stars listed
in Table~\ref{tab:par} are
also available in electronic form at the CDS via anonymous
ftp to cdsarc.u-strasbg.fr (130.79.128.5)
or via http://cdsweb.u-strasbg.fr/Abstract.html
}
}

\author{D.~Montes \inst{1,2}
\and E.L. Mart\'{\i}n \inst{3,4}
}

\offprints{D.~Montes (dmg@astrax.fis.ucm.es)}

\institute{Departamento de Astrof\'{\i}sica,
Facultad de F\'{\i}sicas,
Universidad Complutense de Madrid, E-28040 Madrid, Spain
\and
Pennsylvania State University, Department of Astronomy \& Astrophisics, 
525 Davey Lab, University Park, PA 16802, USA
\and
Instituto de Astrof\'{\i}sica de Canarias, E-38200 La Laguna,
Tenerife, Spain
\and
University of California at Berkeley, Department of Astronomy, 
Berkeley, CA 94720, USA
}

\date{Received ; accepted }
\maketitle

%\markboth{D. Montes et al.:}{ }

%**************************
\begin{abstract}
%**************************

We present a library of Utrecht echelle spectrograph (UES) observations 
of a sample of  F, G, K and M field dwarf stars
covering the spectral range from 4800~\AA$\ $ to 10600~\AA with a 
resolution of 55000.
These spectra include some of
the spectral lines most widely used as
optical and near-infrared indicators of chromospheric activity
such as H$\beta$,  Mg~{\sc i} b triplet,
Na~{\sc i} D$_{1}$, D$_{2}$, He~{\sc i} D$_{3}$,
H$\alpha$, and Ca~{\sc ii} IRT lines, as well as a large number of 
photospheric lines which can also be affected by chromospheric activity.
The spectra have been compiled with the aim of providing a set of 
standards observed at high-resolution to be used in the application of the
spectral subtraction technique to obtain the active-chromosphere
contribution to these lines in
chromospherically active single and binary stars.
This library can also be used for spectral classification purposes.
A digital version with all the spectra is available via ftp and the
World Wide Web (WWW) in both ASCII and FITS formats.


\end{abstract}
%**************************

%**************************
\section{Introduction}
%**************************

Spectral libraries of late-type stars
are a very powerful tool for the study of the chromospheric activity 
by application of the spectral subtraction technique
(see Montes et al. 1995a, b, c; and references therein).
Furthermore, these libraries are also very useful in many  
areas of astrophysics such as the stellar spectral classification, 
modelling stellar atmospheres, stellar abundances, 
calibration of temperatures, spectral synthesis applied to composite systems,  
and spectral synthesis of the stellar population of galaxies.

While for many of this purposes, obtaining as large a spectral range as 
possible was the main priority,
for the chromospheric activity studies, which are centered in 
specific spectral features, it is much  more important to increase the 
spectral resolution.
However, previously published stellar libraries are of poor spectral 
resolution (between 45 and 1.25 \AA) and the only attempt to improve the
spectral resolution is our 
library of high and mid-resolution spectra in the
Ca~{\sc ii} H \& K, H$\alpha$, H$\beta$,
Na~{\sc i} D$_{1}$, D$_{2}$, and He~{\sc i} D$_{3}$
line regions of F, G, K and M field stars
(Montes et al. 1997a, hereafter Paper I)
with resolutions that range between 3 and 0.2 \AA.

However, even more higher resolutions are needed when we are 
interested in very detailed studies of chromospheric activity such as  
the analysis of the difference features present in 
the chromospheric emission line profiles,
the study of chromospherically active binaries aimed 
to determine from which component of the binary belong 
the emission lines (see Montes et al. 1997b),
or the analysis of the time variations and line asymmetries that occur 
during a stellar flare (see Montes et al. 1998).

On the other hand, the simultaneous observations of different lines,
that are formed at different height in the chromosphere
(from the region of temperature minimum to the higher chromosphere),
are of special interest for stellar activity studies since they
provide very useful information about this stellar region.
Ideally, simultaneous observations should be performed at all wavelengths in
order to develop a coherent 3-D atmosphere model
(see the multiwavelength optical observations of 
chromospherically active binary systems by Montes et al. 1997b, c).
So, to carry out these purposes
applying the spectral subtraction technique, to as many lines as possible,
a spectral library with a good spectral resolution 
and a good spectral coverage is needed.

The spectral library that we present in this paper is an extension
of our previous one (Paper I) 
to higher spectral resolution covering a large spectral range.
The library consist of echelle spectra 
of a sample of  F, G, K and M field dwarf stars
covering the spectral range from 4800 \AA$\ $ to 10600 \AA$\ $ and 
with spectral resolution ranging from 0.19 to 0.09 \AA.
These spectra include some of 
the spectral lines most widely used as
optical and near-infrared indicators of chromospheric activity
such as:
Na~{\sc i} D$_{1}$, D$_{2}$, and Mg~{\sc i} b triplet
(formed in the  upper photosphere and lower chromosphere),
Ca~{\sc ii} IRT lines (lower chromosphere),
H$\alpha$, H$\beta$ (middle chromosphere), and
He~{\sc i} D$_{3}$ (upper chromosphere), 
as well as a large number of 
photospheric lines which can also be affected by chromospheric activity. 
Furthermore, the spectra also include a lot of lines of interest to
spectral classification and calibration of temperatures purposes, 
as well as other lines normally used for the application 
of the Doppler imaging technique.

In Sect.~2 we report the details of our observations and data reduction.
The library is presented in Sect.~3 with comments on the behaviour of some
interesting spectral lines.

%******************************************************************
\section{Observations and data reduction}
%******************************************************************

The high-resolution echelle spectra presented here
were obtained by us during several observing runs with the
4.2m William Herschel Telescope (WHT) at La Palma Observatory,
using the Utrecht Echelle Spectrograph (UES) mounted on a Nasmith
focal station.
A description of the WHT/UES is given by Unger (1992).
In addition, we analysed also
different WHT/UES observational campaigns retrieved
from La Palma Data Archive (Zuiderwijk et al. 1994).

In Table \ref{tab:obs} we give a summary of WHT/UES observations,
for each observing run we list the date, the observer, the CCD detector,
 the number of echelle orders included,
the central wavelength ($\lambda$$_{c}$),
the wavelength range covered ($\lambda$$_{i}$-$\lambda$$_{f}$) and the range 
of reciprocal  dispersion  achieved (\AA/pixel) from the first to the last 
echelle orders. The echelle grating  used in all these runs was the E31. 
The interorder spacing was too small to perform a good sky subtraction. 
Sky contamination is only significant for the few faint very-late M-type 
stars of this library (VB~8 and VB~10). 
As can be seen in Table~1, the spectra cover
the spectral range from 4800 \AA$\ $ to 10600 \AA$\ $ 
with spectral resolution (FWHM=2 pixels) ranging from 0.19 to 0.09 \AA, 
corresponding to R$\sim$55000. 
However, as the CCD chips we have used were smaller than
the echelle orders, we did not get a full recording of the different orders.
The gaps between the adjacent recorded orders got larger towards longer 
wavelengths, as can be seen in 
Tables~\ref{tab:jul93} to \ref{tab:nov94},
 where we give for each observing run the wavelength range and the
spectral lines of interest in each echelle order.

The spectra have been extracted using the standard
reduction procedures in the IRAF
\footnote{IRAF is distributed by the National Optical Observatory,
which is operated by the Association of Universities for Research in
Astronomy, Inc., under contract with the National Science Foundation.}
 package (bias subtraction,
flat-field division, and optimal extraction of the spectra).
The wavelength calibration was obtained by taking
spectra of a Th-Ar.
Finally, the spectra have been normalized by
a polynomial fit to the observed continuum.

%****************************************************************
\section{The library}
%****************************************************************

As in Paper I, the stars included in the library
have been selected as stars with low levels of chromospheric activity,
that is to say, 
stars that do not present any evidence of emission in the core
of Ca~{\sc ii} H \& K lines in our spectra 
(Montes et al. 1995c, 1996a),
stars with the lower Ca~{\sc ii} H \& K spectrophometric index S
(Baliunas et al. 1995),
or stars known to be inactive and slowly rotating stars 
from other sources 
(see Strassmeier et al. 1990; Strassmeier \& Fekel 1990; Hall \& Ramsey 1992).
In addition, we provide spectra of some active stars of late and very-late 
spectral types.

Table~\ref{tab:par} presents information about the observed stars.
In this table we give the HD, HR and  GJ numbers, name, 
spectral type and luminosity class (T$_{\rm sp}$),
from the Bright Star Catalogue
(Hoffleit \& Jaschek 1982; Hoffleit \& Warren 1991) and 
the Catalogue of Nearby Stars (Gliese \& Jahreiss 1991), 
except for some M dwarfs for which we list the more recent 
spectral type determination given by 
Henry et al. (1994) and Kirkpatrick et al. (1995).
In column (6) MK indicates if the star is included in the list of
Morgan and Keenan (MK) Standard Stars compiled by Garc\'\i a (1989).
Column (7) give the metallicity [Fe/H] from Taylor (1994; 1995)
or Cayrel de Strobel (1992; 1997)
 and  column (8) rotational period (P$_{\rm rot}$) and {\it v}~sin{\it i}
from Donahue (1993), Baliunas et al. (1995),
%The values of {\it v}~sin{\it i} marked with "*" and  "F" 
%are from the references given in 
Strassmeier \& Fekel (1990), Fekel (1997), Stauffer \& Hartmann (1986),
Mart\'{\i}n et al. (1996), and Basri \& Marcy (1995; 1996). 
We also give, in column (9), the Ca~{\sc ii} H \& K spectrophometric index S
from Baliunas et al. (1995) and  Duncan et al. (1991).
In column (10) we list information about the observing run in which each 
star have been observed, using a code given
in the first column of  Table~\ref{tab:obs}, and 
the last column indicate if the star was also included in Paper I.

In Fig.~\ref{fig:orders} we have plotted for a K1V star
representative spectral orders,
with the line identification marked.
The two first orders (H$\beta$ and Mg~{\sc i} b lines) 
correspond to the K1V star HD 10476 from the Dec-93 observing run and
the following orders correspond to K1V star HD 9546 from the the Nov-94 run.

%In Fig.~\ref{fig:orders} we have plotted, 
%for one observing run ( - ), 
%all the spectral orders, 
%except which are strongly contaminated by telluric absorption bands,
%of one of the object (a K0V star) with the line identification marked.

Representative spectra (from F to M stars) 
in different spectral regions are plotted 
in figures (\ref{fig:hb} to \ref{fig:cairt})
in order to show the behaviour
of the more remarkable spectroscopic features
with the spectral type.
In order of increasing wavelength we have plotted
the following line regions:
H$\beta$ (Fig.~\ref{fig:hb}), 
Mg~{\sc i} b triplet (Fig.~\ref{fig:mg})
Na~{\sc i} D$_{1}$, D$_{2}$ (Fig.~\ref{fig:na}) and  
He~{\sc i} D$_{3}$) (Fig.~\ref{fig:na}),
H$\alpha$ (Fig.~\ref{fig:ha}),and 
Ca~{\sc ii} IRT $\lambda$8498, 8542, 8662  (Fig.~\ref{fig:cairt}).

In the following,  we describe the behaviour of some 
interesting spectral lines and molecular bands 
present in the spectral range covered by the spectra (from 4800 to 10600 \AA).
We list the spectroscopic features in order of increasing wavelength, and 
the echelle order in which they appear in each observing run can be found
in Tables~\ref{tab:jul93} to \ref{tab:nov94}. 
%, \ref{tab:dec93}, \ref{tab:apr94} 
%and  \ref{tab:nov94}.

%.....................................
% Spectral lines of interest
%.....................................
\begin{itemize}

\item The H$\beta$ $\lambda$4861.3 line (Fig.~\ref{fig:hb})
is a well know chromospherically activity indicator (emission or filled-in).

%\item The Fe~{\sc i} $\lambda$4921 line.

\item The Mg~{\sc i} b triplet  $\lambda$$\lambda$5167, 5172, 5183 
(Fig.~\ref{fig:mg}) is luminosity sensitive in the range G8-K5.
These strong neutral metal lines  are formed in the lower chromosphere and the
region of temperature minimum and they are good diagnostics of
activity (Basri et al. 1989; Gunn \& Doyle 1997; Gunn et al. 1997;
Montes et al. 1998).

\item The He~{\sc i} D$_{3}$ $\lambda$5876 absorption line 
is another important indicator of stellar activity in the upper
chromosphere (Garc\'{\i}a L\'{o}pez et al. 1993; Montes et al. 1997b, c;
Saar et al. 1997) and also could be in emission during stellar flares
(Huenemoerder \& Ramsey 1987;
Montes et al. 1996b; 1997b; 1998).

\item The Na~{\sc i} D$_{1}$ $\lambda$5895.92 and D$_{2}$~$\lambda$5889.95 
lines (Fig.~\ref{fig:na})
are well known temperature and luminosity discriminant.
These resonance lines are collisionally-controlled in the
atmospheres of late-type stars and then provide information about 
chromospheric activity see Montes et al. (1996b, 1997b, c) and the recent
models of these lines for M dwarfs stars by Andretta et al. (1997).

\item The wings of the Ca~{\sc i} triplet $\lambda$$\lambda$6102, 6122, 6162 
lines can be used as luminosity indicators (Cayrel et al. 1996). 
These lines are very weak
at spectral type F and increase in strength with decreasing temperature.

\item The V~{\sc i} $\lambda$6251.83 and Fe~{\sc i}  $\lambda$6252.57 
line-depth ratio can be used to determine stellar temperatures 
(Gray \& Johanson 1991, Gray 1994). 

\item The line ratio Fe~{\sc ii}~$\lambda$6432.65/Fe~{\sc i}~$\lambda$6430.85
is useful for spectral-class/temperature classification for F to M stars.
Other spectral class indicators are the ratios of
V~{\sc i}~$\lambda$6452/Ca~{\sc i}~$\lambda$6456 (for F, and G stars),
Co~{\sc i}~$\lambda$6455/Ca~{\sc i}~$\lambda$6456 and
Fe~{\sc ii}~$\lambda$6457/Ca~{\sc i}~$\lambda$6456 (for F to K stars) 
(Strassmeier \& Fekel 1990).

\item The Fe~{\sc i} 6411.66~\AA, Fe~{\sc i} 6430.85~\AA, 
and Ca~{\sc i} 6439.08~\AA$\ $ lines
normally used for the application of the Doppler imaging technique.

%\item The Fe~{\sc i} 6495~\AA, 6546.25~\AA, 6663.4~\AA, 
%and 6677.9~\AA$\ $ lines and the Ca~{\sc i} 6718~\AA$\ $ line
%are strong absorption lines that
%could be used for radial and rotational velocity determinations,
%and its intensity increases toward later spectral types.

\item The emission or filled-in of the
H$\alpha$~(6562.8~\AA) (Fig.~\ref{fig:ha})
line is one the primary optical 
indicators of chromospheric activity
in late-type stars, together with the 
Ca~{\sc ii} H \& K emission lines.

%\item The Ca~{\sc i} 6572.78~\AA$\ $ resonance line.

\item The Li~{\sc i} resonance line at $\lambda$6707.8~\AA$\ $  
and the subordinate lines at $\lambda$6103~\AA$\ $ 
and $\lambda$8126~\AA$\ $ 
(Pavlenko et al. 1995; Carlsson et al. 1994).

\item  The K~{\sc i} $\lambda$7664.91, 7698.98 doublet is a
resonance transition and thus
will be influenced by chromospheric activity, 
particularly in M dwarf stars 
(see Basri \& Marcy 1995; Schweitzer et al. 1996).

%\item The Fe~{\sc i} 7834~\AA$\ $ line observed in K dwarfs.

%\item The Fe~{\sc i} 7913~\AA$\ $ and  Fe~{\sc i} 7937~\AA$\ $ lines.

\item The Rb~{\sc i} resonance line at 7947.63 ~\AA$\ $
is strong in very late-type stars, and it seems to be a good temperature
diagnostic (Basri \& Marcy 1995).

\item The Na~{\sc i} doublet $\lambda$$\lambda$8183.26, 8194.8 lines are
subordinate transitions and therefore, form mainly in the photosphere
and should not be significantly affected by the chromosphere.
These lines can be used as a dwarf/giant indicator (Schiavon et al. 1997).

\item The Ti~{\sc i} lines for multiplet 33 
(specially the $\lambda$8382.54, 8382.82 lines) and
other Ti~{\sc i} lines.
% such as
%$\lambda$5366.651 (Multiplet 35), $\lambda$5446.593 (M. 3),
%$\lambda$5866.453, $\lambda$5937.806 and $\lambda$5941.755 (M. 72),
%$\lambda$6554.226 and $\lambda$6556.066 (M.102), and $\lambda$6599.112 (M.19), 
can be used as a luminosity classification criterium
because they present a positive luminosity effect
(Keenan \& Hynek 1945; 
Ginestet et al. 1994;  Jashek \& Jashek 1995; Montes et al. 1997c).

\item The Ca~{\sc ii} infrared triplet (IRT) $\lambda$$\lambda$8498, 8542, 
and 8662 lines (Fig.~\ref{fig:cairt})
formed in the lower chromosphere
and are also important activity indicators
(Linsky et al. 1979; Foing et al. 1989; Dempsey et al. 1993;
Montes et al. 1997c). 
These lines have been also used by several authors 
as a gravity sensitive index.
In this spectral region the Paschen lines P12 ($\lambda$8750) and 
P14 ($\lambda$8598) are also visible in F stars
(Carquillat et al. 1997).

%\item The Mg~{\sc i} 8807~\AA$\ $ line.

\item From mid K through M stars we can also see a number of 
titanium oxide (TiO) molecular bands such as 
(5847-6058), (6080-6390), (6322-6512), (6569-6649), (6651-6852), 
(7053-7270), (7666-7861), (8206-8569), (8432-8452), and  (8859-8937), 
useful in classifying early M dwarfs, 
and a number of vanadium oxide (VO) bands such as 
(7400-7510), (7851-7973), and (8521-8668), 
useful in classifying late M dwarfs.
CaH bands at (6346, 6382, 6389) and  (6750-7050) are also present
in these stars.

\item Other strong features that appear in the spectra are the
telluric O$_{2}$ A and B bands at 6867~\AA$\ $ 7600 \AA$\ $ and 
several telluric H$_{2}$O bands at (7186-7273), (8164-8177), 
8227, 8282, (8952-8972), and (8980-8992).

\end{itemize}

%...................................................................

A more detailed description of the
H$\alpha$, H$\beta$, 
Na~{\sc i} D$_{1}$, D$_{2}$ and He~{\sc i} D$_{3}$ lines 
and the corresponding photospheric features included in these spectral 
regions, can be found in Paper I.
An extensive list of features identifiable in late-K to late-M spectra
from 6300 to 9000 \AA$\ $ can be found in Kirkpatrick et al. (1991).
A description of the spectroscopic characteristics of very cool dwarfs
and substellar candidates is given by  Mart\'{\i}n et al. (1996).
For more information about spectral classification of stars and 
the behaviour of chemical elements in stars the reader is referred to
Jaschek \& Jaschek (1990; 1995).

In order to enable other investigators to make use of the spectra of this
library, all the multidimensional spectra containing all the echelle orders
of the stars listed
in Table~\ref{tab:par} are available
as FITS format files at the CDS in Strasbourg, France,
via anonymous ftp to cdsarc.u-strasbg.fr (130.79.128.5).
They are also available via the World Wide Web at:
\newline
{\small http://www.ucm.es/OTROS/Astrof/fgkmsl/UESfgkmsl.html}.

In order to facilitate the use of this library 
the onedimensional normalized and wavelength-shifted 
spectra, resulting for the extraction of the orders
containing the more remarkable spectroscopic 
features (the spectral regions plotted in 
figures~\ref{fig:hb} to \ref{fig:cairt}), 
are also available as separate FITS format files.

The extension of this library including stars of higher luminosity class, 
as well as the use of these spectra to analyse temperature sensitive lines
in order to improve the actual line-depth ratio temperature calibrations
(Gray \& Johanson 1991, Gray 1994) 
and spectral-class/temperature classifications 
(Strassmeier \& Fekel 1990),
will be the subject of forthcoming papers.

%---- Tables  ----------------------------------------------------

%..................................................................
\begin{table}
\caption[ ]{Summary of WHT/UES observations 
(published only electronically at CDS) 
\label{tab:obs} }
\end{table}
%............................................

%******* Table  ***********************************************
\begin{table}
\caption[ ]{UES spectral orders (Jul-93)
(published only electronically at CDS)
\label{tab:jul93} }
\end{table}                                           
%**************************************************************

%******* Table  ***********************************************
\begin{table}
\caption[ ]{UES spectral orders (Dec-93)
(published only electronically at CDS)
\label{tab:dec93} }
\end{table}                                           
%**************************************************************

%******* Table  ***********************************************
\begin{table}
\caption[ ]{UES spectral orders (Apr-94, Mar-95, Jun-95)
(published only electronically at CDS)
\label{tab:apr94} }
\end{table}
%**************************************************************

%******* Table  ***********************************************
\begin{table}
\caption[ ]{UES spectral orders (Nov-94)
(published only electronically at CDS)
\label{tab:nov94} }
\end{table}
%**************************************************************

%%+++++++++++++++++++
\begin{figure*}
\caption[ ]{Representative spectral orders for the Dec-93 and Nov-94 
observing runs of a K1V star,  with the line identification marked
(published only electronically at CDS)
\label{fig:orders}}
\end{figure*}
%%+++++++++++++++++++

%******* Table ********************************************************
\begin{table*}
\caption[ ]{Stars
\label{tab:par} }
\begin{flushleft}
%\begin{center}
\scriptsize
%\small
%---------------1 2 3 4 5 8 9 10 11
\begin{tabular}{l l l l l l l l l l l l l}
\hline
\noalign{\smallskip}
%.........................................................................
 HD     & HR   & GJ    &   Name        & T$_{\rm sp}$  & MK &
 [Fe/H] & P$_{\rm rot}$ & {\it v}~sin{\it i} & S & Obs. & Pap. I \\
        &      &       &               &               &    & 
 (dex)  & (days)        & (km s$^{-1}$)      &   &      &        \\
%.........................................................................
\noalign{\smallskip}
\hline
 {\bf F stars}  \\
\hline
\noalign{\smallskip}
33256   & 1673 & 189.2 & 68 Eri        & F2V           &    & 
-0.49   & -    & 0   & -    & 3 &  \\ 
6406    &      &       &               & F3/F5IV/V     &    & 
 -    & -    & -   & -    & 3 &  \\
84937   &      &       & BD+14 2151    & F5VI          &    &
-1.86  & -    & -   & -    & 2 &  \\
        &      &       & BD+18 3423    & F6V           &    &
 -    & -    & -   & -    & 1 &  \\
78154   & 3616 &    & $\sigma^{2}$ UMa & F6IV          &    &
 -    & -    & 0   & 0.142 & 6 &  \\
9826    & 458  & 61    & 50 And        & F8V           &  MK& 
-0.14   & -    & 8   & 0.154 & 1 &  \\
142373  & 5914 & 602   & $\chi$ Her    & F8V           &    & 
-0.431& -    & 10.0  & 0.147 & 5 & *\\
144284  & 5986 & 609.1 & $\theta$ Dra  & F8IV  (SB1)    &    & 
0.23   & -    & 27.7   & 0.203 & 6   & \\
98230   & 4374 & 423B  & $\xi$ UMa B   & F8.5V         &  MK & 
-0.12  & -    & 3   & -    & 5 &   \\
114762  &      &       & BD+18 2700    & F9V           &    & 
-0.87  & -    & -   & -    & 3 &   \\  
79028   & 3648 & 337.1 & 16 UMa        & F9V  (SB1)     &    & 
 -    & -    & 0   & -    & 5, 6 &   \\
114710  & 4983 & 502   & $\beta$ Com   & F9.5V         &  MK&
0.135& 12.35 &4.3F& 0.201 & 3 & * \\  
\noalign{\smallskip}
\hline
{\bf G stars} \\
\hline
\noalign{\smallskip}
160269  & 6573 &       & 26 Dra        & G0IV-V (MK) (SB1) &  MK& 
 -    & -    & 41   & -    & 1 &  \\
15335   & 720  & 99.1  & 13 Tri        & G0V           &    & 
 -    & -    & $<$ 6 & -    & 4 &  \\
39587   & 2047 & 222B  & $\chi^{1}$ Ori& G0V  (SB1)    &  MK& 
-0.084& 5.36  & 8.6F   & 0.325 & 3 & * \\ 
98231  A& 4375 & 423A  & $\xi$ UMa A   & G0V (SB1)     &    & 
-0.352& -     & $<$  15& -     & 5 & * \\
84737   & 3881 &       & 15 LMi        & G0.5V         &  MK& 
-0.04  & -    & 3   & 0.145 & 5 &  \\
10307   & 483  & 67    & BD+41 328     & G1.5V  (SB1)  &  MK& 
0.14    & -    & 2.1   & 0.152 & 2 &  \\
42807   & 2208 & 230   & BD+10 1050    & G2V           &    & 
 -    & -    & -   & 0.352 & 4 &  \\
153631  &      & 650   & BD-13 4528    & G2V  (SB1)    &    &
 -    & -    & -   & -    & 3 &   \\
186427  & 7504 & 765.1B& 16 Cyg B      & G3V           &  MK& 
-0.002 & -     &  0.4   & 0.145 & 1 & * \\
86728   & 3951 & 376   & 20 LMi        & G3V           &  MK& 
-0.11     & -    & 3   & 0.156 & 6 &  \\ 
%82210   & 3771 & 355.1 & DK UMa        & G4III-IV (G5III-IV MK)    &  MK&
% -    & -    & $<$  19   & 0.398  & 2 &   \\
115617  & 5019 & 506   & 61 Vir        & G5V           &    & 
 0.032& -     & 0.4   & 0.162 & 3 & * \\  
178428  & 7260 & 746   & BD+16 3752    & G5V  (SB1)     &    & 
 -   & -    & -   & 0.154 & 3, 6  &  \\
33802 B &      &       & BD-12 1095B   & G5Ve          &    & 
 -    & -    & -   & -    & 3  &  \\
149414  &      & 629.2A& BD-03 3968    & G5V (SB1)      &    & 
-1.14  & -    & -   & -    & 3, 5, 6   &  \\
20630   & 996  & 137   &$\kappa^{1}$ Cet&G5V           &  MK& 
0.133& 9.24 & 3.9    & 0.366 & 2 & *  \\
31966   &      & 182.1 & BD+14 804     & G5V           &    & 
 -    & -    & -   & -    & 4 &  \\
108754  &      & 469.1 & BD-02 3528    & G7V  (SB1)    &    & 
 -    & -    & -   & -    & 3, 5 & \\
131156 A&5544 A& 566A  & $\xi$ Boo A   & G8V           &  MK& 
-0.151& 6.31  & 3.2   & 0.461 & 5 & *  \\
44867   & 2302 &       & BD+16 1135    & G8IV (G9III)  &    &
 -    & -    & -   & -    & 4 &  \\
195987  &      & 793.1 & BD+41 3799    & G9V  (SB1)     &    & 
 -    & -    & -   & -    & 3 &  \\
\noalign{\smallskip}
\hline
{\bf K stars} \\
\hline
\noalign{\smallskip}
10780   & 511  & 75    & BD+63 238     & K0V           &    & 
0.36  & -     & 0.6  & 0.280 & 4 & \\
185144  & 7462 & 764   & $\sigma$ Dra  & K0V           &  MK& 
-0.045& -      & 0.6  & 0.215 & 1 & * \\
18972   &      &       & BD+13 494     & K0IV          &    & 
 -    & -    & -   & -    & 4 &  \\
48432   & 2477 &       & 13 Lyn        & K0III         &    & 
 -    & -    & $<$  19 & 0.120  & 2 &  \\ 
9546    &      & 59.3  & ADS 1233 A    & K1V           &    & 
 -    & -    & -   & -    & 4 &  \\
10476   & 493  & 68    & 107 Psc       & K1V           &  MK& 
-0.123& 35.2 &  0. & 0.198 & 2 & * \\
76291   & 3545 &       & BD+46 1459    & K1IV          &    & 
 -    & -    & -   & -    & 2 &  \\
6027    &      & 50.1  & BD+58 155     & K2V           &    & 
 -    & -    & -   & -    & 4 &  \\
101177 B& 4486 B&433.2 B& ADS 8250 A   & K2V (SB)      &    & 
 -    & -    & -   & 0.144  & 3, 6  &  \\
38392   & 1982 & 216B  & $\gamma$ Lep B& K2V           &    & 
0.02     & -    & -   & -    & 3 &  \\ 
136713  &      & 1191  & BD-10 4088    & K2V           &    & 
 -    & -    & -   & -    & 3 &  \\
223778  & 9038 & 909 A & BD+74 1047    & K3V           &    & 
 -    & -    & -   & -    & 1 &  \\
219134  & 8832 & 892   & BD+56 2966    & K3V           &  MK& 
-9.000 & -      & -     & 0.230 & 2 & * \\
16160 A &  753 & 105A  & BD+06 398     & K3V           &  MK& 
-0.297 & 48.0   & -     & 0.226 & 4 & * \\ 
98800   &      & 2084A & ADS 8141 A    & K4V           &    & 
 -    & -    & -   & -    & 3, 5 &  \\
131156 B&5544 B& 566B  & $\xi$ Boo B   & K4V           &    & 
0.19   & 12.28  & 20    & 1.381 & 5 & * \\
12208   &      & 83.3  & V598 Cas      & K5V           &    & 
 -    & -    & -   & -    & 4 &  \\
154363  &      & 653   & BD-04 4225    & K5V           &    & 
 -    & -    & -   & -    & 3 &  \\
201091  & 8085 & 820 A & 61 Cyg A      & K5V           &  MK& 
-0.06  & 35.37  & 0.6    & 0.658 & 2, 6 & * \\
201092  & 8086 & 820 B & 61 Cyg B      & K7V           &  MK& 
-0.10  & 37.84  & 1.4    & 0.986 & 2, 6 & * \\
        &      & 52    & BD +63 137    & K7V           &    & 
 -    & -    & -   & -    & 4 &   \\
157881  &      & 673   & BD+02 3312    & K7V           &    & 
0.40   & -    & 3.9   & 1.464 & 1 &   \\
88230   &      & 380   & BD+50 1725    & K7V(1) (K6V MK)& MK& 
0.28   & -    & 3.1  & 1.617 & 3 &   \\
151877  &      & 639   & BD+37 2804    & K7V           &    & 
 -    & -    & -   & 0.197 & 3 &   \\
151288  &      & 638   & BD+33 2777    & K7.5Ve        &  MK& 
 -    & -    & -   & 1.380 & 3, 6 &  \\
\noalign{\smallskip}
\hline
\end{tabular}
\end{flushleft}
%\end{center}
\end{table*}

%******* Table 1 cont ***********************************************
\begin{table*}
\addtocounter{table}{-1}
\caption[ ]{Continue}
\begin{flushleft}
%\begin{center}
\scriptsize
%\small
%---------------1 2 3 4 5 8 9 10 11
\begin{tabular}{l l l l l l l l l l l l l}
\hline
\noalign{\smallskip}
%.........................................................................
 HD     & HR   & GJ    &   Name        & T$_{\rm sp}$  & MK &
 [Fe/H] & P$_{\rm rot}$ & {\it v}~sin{\it i} & S & Obs. & Pap. I \\ 
        &      &       &               &               &    & 
 (dex)  & (days)        & (km s$^{-1}$)      &   &      &        \\
%.........................................................................
\noalign{\smallskip}
\hline
{\bf M stars} \\
\hline
\noalign{\smallskip}
        &      & 16    &               & M0V           &    & 
 -    & -    & -   & -    & 1 &  \\  
79210   &      & 338A  & ADS 7251 A    & M0Ve (1)      &    & 
 -    & -     &  -     & 2.113 & 3 & * \\    
        &      & 572   & BD+45 2247    & M0V           &    & 
 -    & -    & -   & -    & 3, 5 &  \\
232979  &      & 172   & BD+52 857     & M0.5V         &  MK& 
 -    & -    & -   & 1.909  & 4 &  \\
1326 A  &      & 15A   & GX And        & M1.5V (1) (M2V MK)& MK& 
 -    & -    & -   & -    & 4 &  \\
36395   &      & 205   & BD-03 1123    & M1.5V (1)     &  MK& 
0.60   & -    & -   & -    & 4 &  \\
95735   &      & 411   & BD+36 2147    & M2V           &  MK& 
-0.20  & -    & -   & 0.424  & 3  & \\ 
        &      & 623AB & LHS 417       & M2.5V (1)     &    & 
 -    & -    & -   & -    & 3, 5, 6 &  \\ 
        &      & 813   & LHS 3605      & M3V           &    & 
 -    & -    & -   & -    & 1 &  \\
173739 A&      & 725A  & ADS 11632 A   & M3V (1)       &    & 
 -    & -    & -   & 0.534 & 3, 5 & \\
180617  &      & 752A  & LHS 473       & M3 V (1)      &  MK& 
 -    & -    & -   & 1.252  & 6 &  \\
        &      & 273   & BD+05 1668    & M3.5V (1)     &    & 
 -    & -    & -   & -    & 3 &  \\
16160 B &      & 105B  & BD+06 398 B   & M3.5V (1) (3) &    & 
 -    & -    & -   & -    & 1 &  \\
        &      & 699   & Barnard's star& M4V (1)       &    & 
 -    & -    &  -      & -    & 3 &   \\
13124        &      & 748   & Wolf 1062     & M4V (M3.5V (4))&   & 
 -    & -    & -   & -    & 3, 5, 6 &  \\  
        &      & 447   & FI Vir, LHS 315&M4V (1) (3)   &    & 
 -    & -    & -   & -    & 6 &  \\
12025   &      &       & U Per         & M4III (M6e)   &    & 
 -    & -    & -   & -    & 4 &  \\
        &      & 234AB & LHS 1849/50   & M4.5V (1) e   &    & 
 -    & -    & $<$  10   & -    & 3, 5 & \\
        &      & 831AB & LHS 511       & M4.5V (1) e   &    & 
 -    & -    & $<$  10   & -    & 6 &  \\
        &      & 473AB &FL Vir, LHS 333& M5.5V (1) e   &    & 
 -    & -    & -   & -    & 3, 5, 6 &  \\
        &      & 1245A & V1581 Cyg     & M5.5V e       &    & 
 -    & -    & -   & -    & 6 &  \\
        &      & 1245B & LHS 3495      & M5.5V e       &    & 
 -    & -    & -   & -    & 6 &  \\
        &      & 406   & LHS 36        & M6V (1)  e    &    & 
 -    & -    & $<$ 3 & -    & 3 & * \\
        &      & 1111  & DX Cnc, LHS 248&M6.5V         &    & 
 -    & -    & 11   & -    & 3, 5 & * \\
        &      & 644C  & VB 8, LHS 429 & M7V (2)       &    & 
 -    & 0.14  & 8  & -    & 3, 5, 6 &  \\
        &      & 752B  & VB 10, V1298 Aql & M8V (1) e  &    & 
 -    & -    & $<$ 5  & -    & 6 &  \\
%
%.........................................................................
%
\noalign{\smallskip}
\hline
\end{tabular}
\end{flushleft}
%\end{center}

{\scriptsize
(1): Henry et al. (1994)

(2): Kirkpatrick et al. (1995) 

(3): "Zero H$\alpha$ star", Byrne (1993)

(4): Kirkpatrick et al. (1991)

MK: "A List of MK Standard Stars", 
Garc\'{\i}a  (1989)

SB: Spectroscopic Binary (Duquennoy \& Mayor; Mazeh et al. 1997)

%e: emission
}

\end{table*}                                           

%*********************************************************************

%********************************************************
\begin{acknowledgements}
We thank J. Sanz-Forcada for help in the reduction
of some of the echelle spectra.
This research has made use the La Palma Archive and 
of the SIMBAD data base, operated at CDS,
Strasbourg, France.
This work has been supported by the Universidad Complutense de Madrid
and the Spanish Direcci\'{o}n General de Investigaci\'{o}n
Cient\'{\i}fica y  T\'{e}cnica (DGICYT) under grants PB94-0263 
and PB95-1132-C02-01.

\end{acknowledgements}

%*******************************************************************

%********************************************************

%\newpage

%---- Figures ----------------------------------------------------

%%+++++++++++++++++++
\begin{figure*}
{\psfig{figure=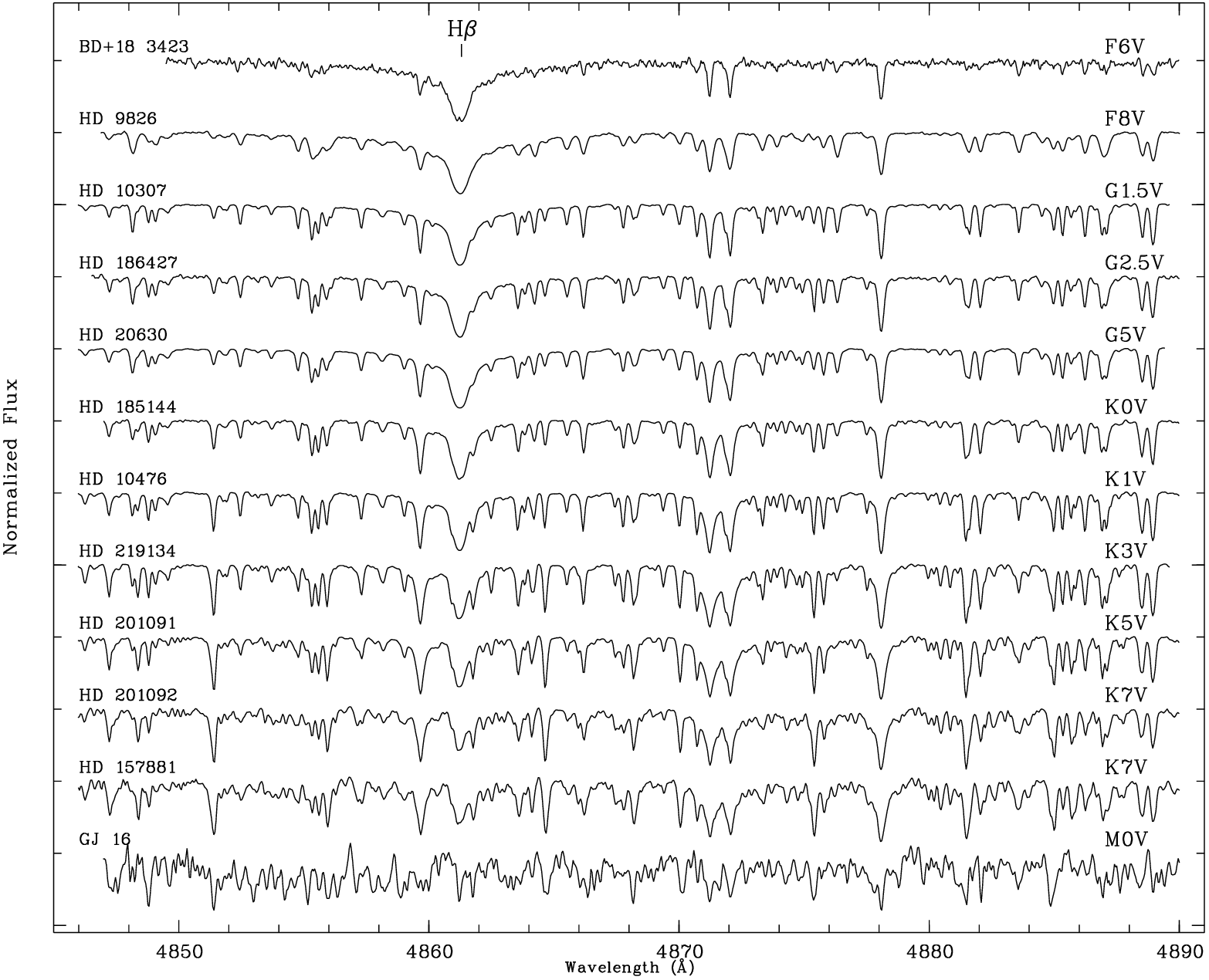,bbllx=34pt,bblly=164pt,bburx=665pt,bbury=678pt,height=22.4cm,width=17.8cm,clip=}}
\caption[ ]{Spectra in the H$\beta$ line region
\label{fig:hb}}
\end{figure*}
%%+++++++++++++++++++

%%+++++++++++++++++++
\begin{figure*}
{\psfig{figure=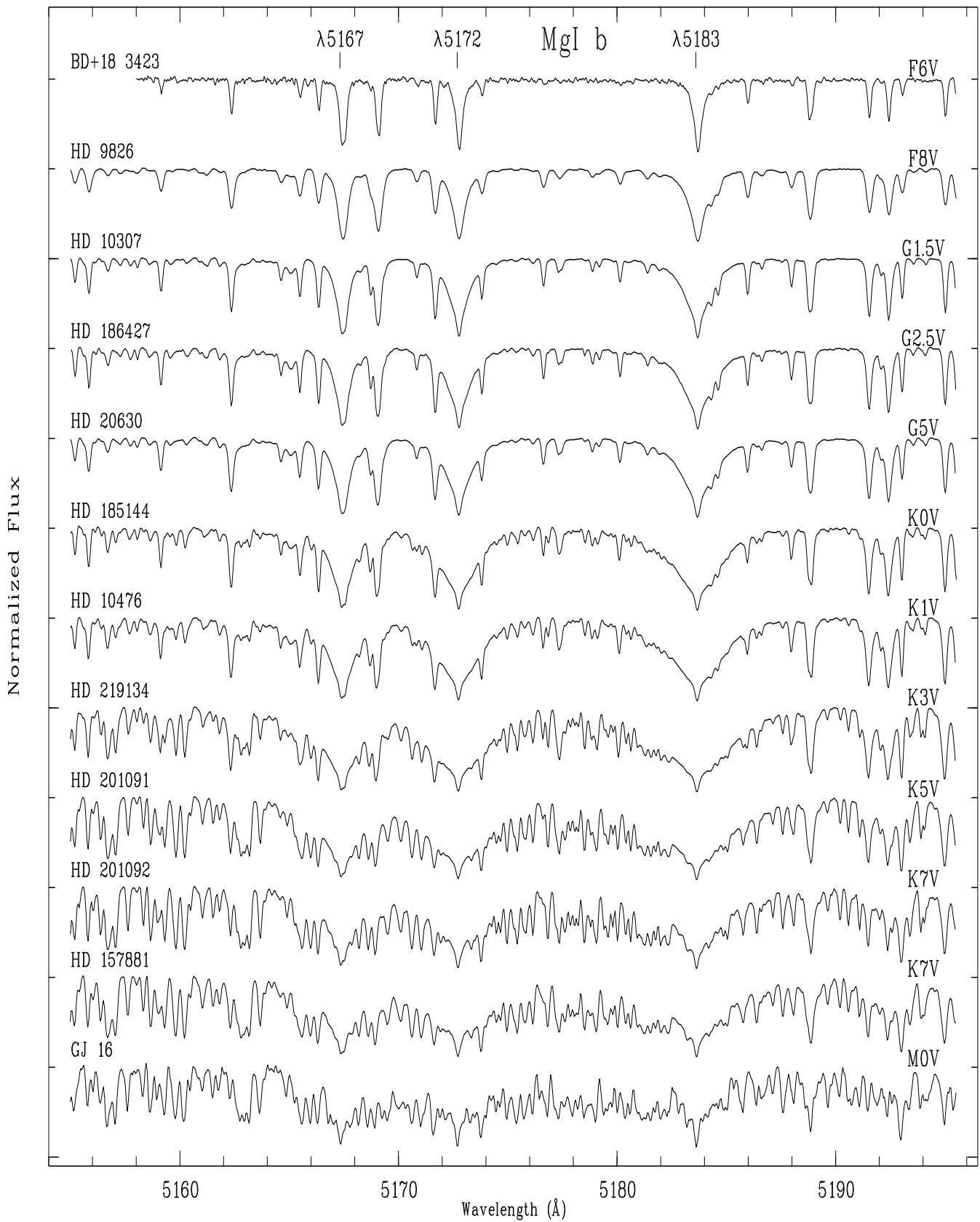,bbllx=34pt,bblly=164pt,bburx=665pt,bbury=678pt,height=22.4cm,width=17.8cm,clip=}}
\caption[ ]{Spectra in the Mg~{\sc i} b triplet lines region
\label{fig:mg}}
\end{figure*}
%%+++++++++++++++++++

%%+++++++++++++++++++
\begin{figure*}
{\psfig{figure=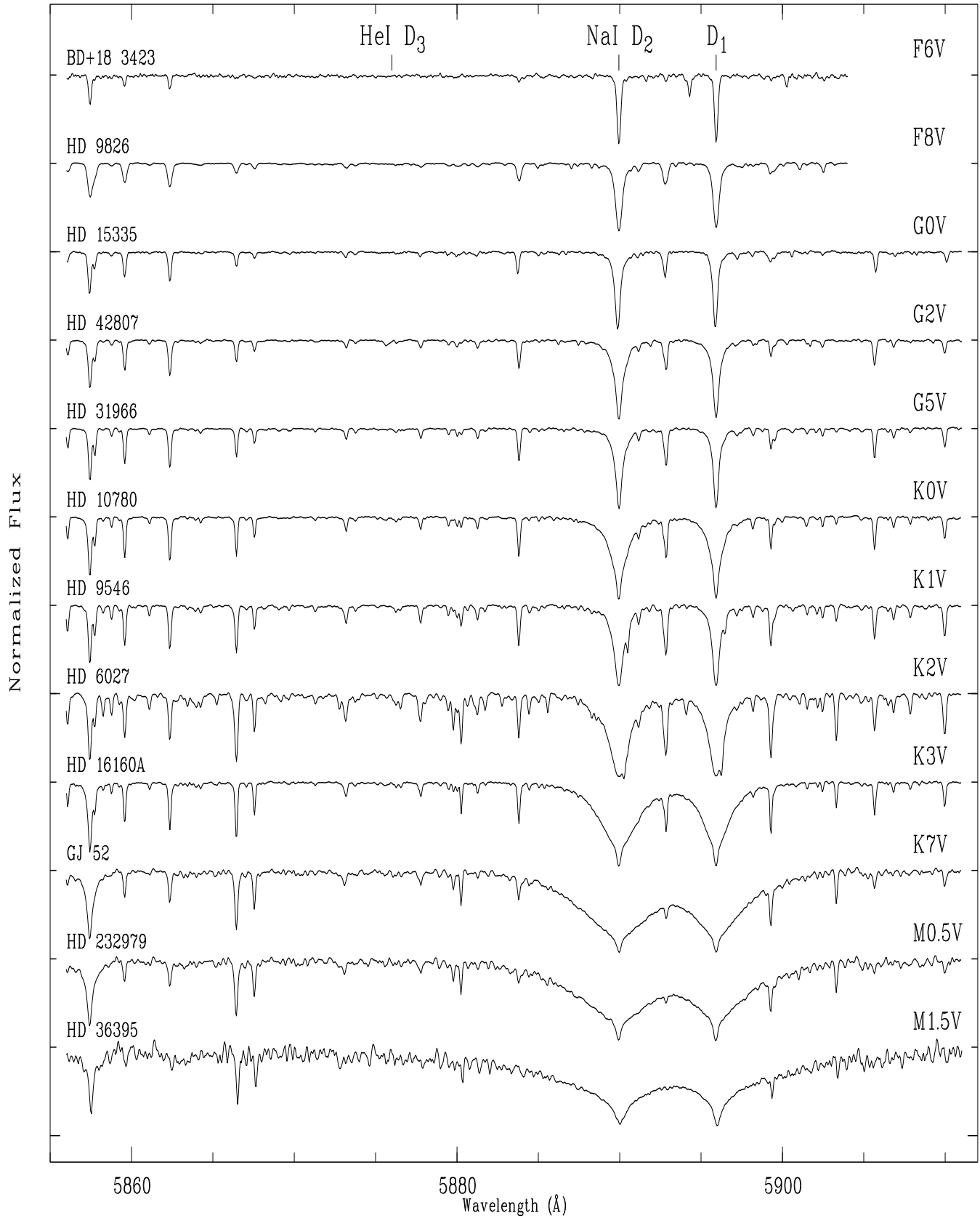,bbllx=34pt,bblly=164pt,bburx=665pt,bbury=678pt,height=22.4cm,width=17.8cm,clip=}}
\caption[ ]{Spectra in the  Na~{\sc i} D$_{1}$, D$_{2}$ and
He~{\sc i} D$_{3}$  lines region
\label{fig:na}}
\end{figure*}
%%+++++++++++++++++++

%%+++++++++++++++++++
\begin{figure*}
{\psfig{figure=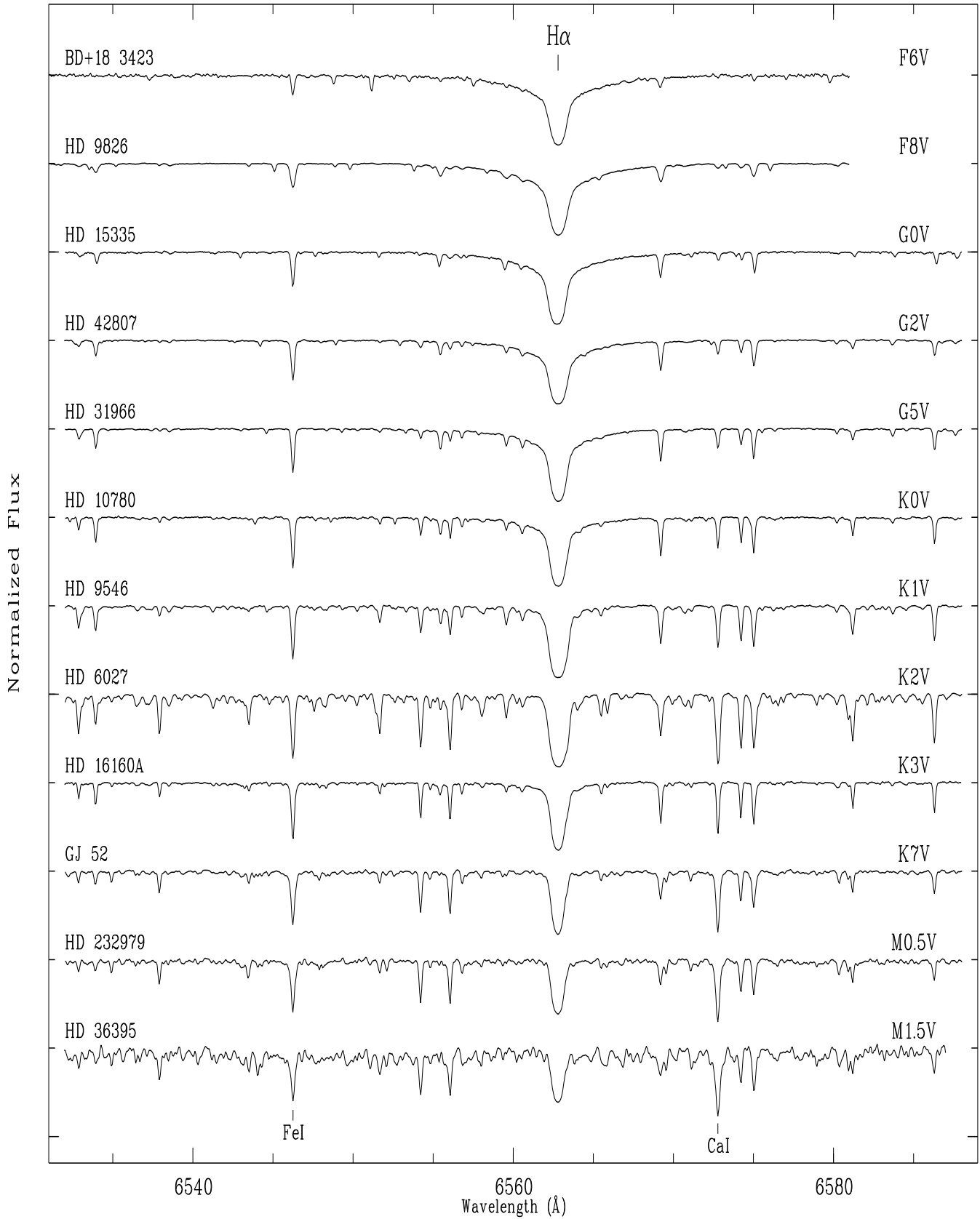,bbllx=34pt,bblly=164pt,bburx=665pt,bbury=678pt,height=22.4cm,width=17.8cm,clip=}}
\caption[ ]{Spectra in the H$\alpha$ line region
\label{fig:ha}}
\end{figure*}
%%+++++++++++++++++++

%%+++++++++++++++++++
\begin{figure*}
{\psfig{figure=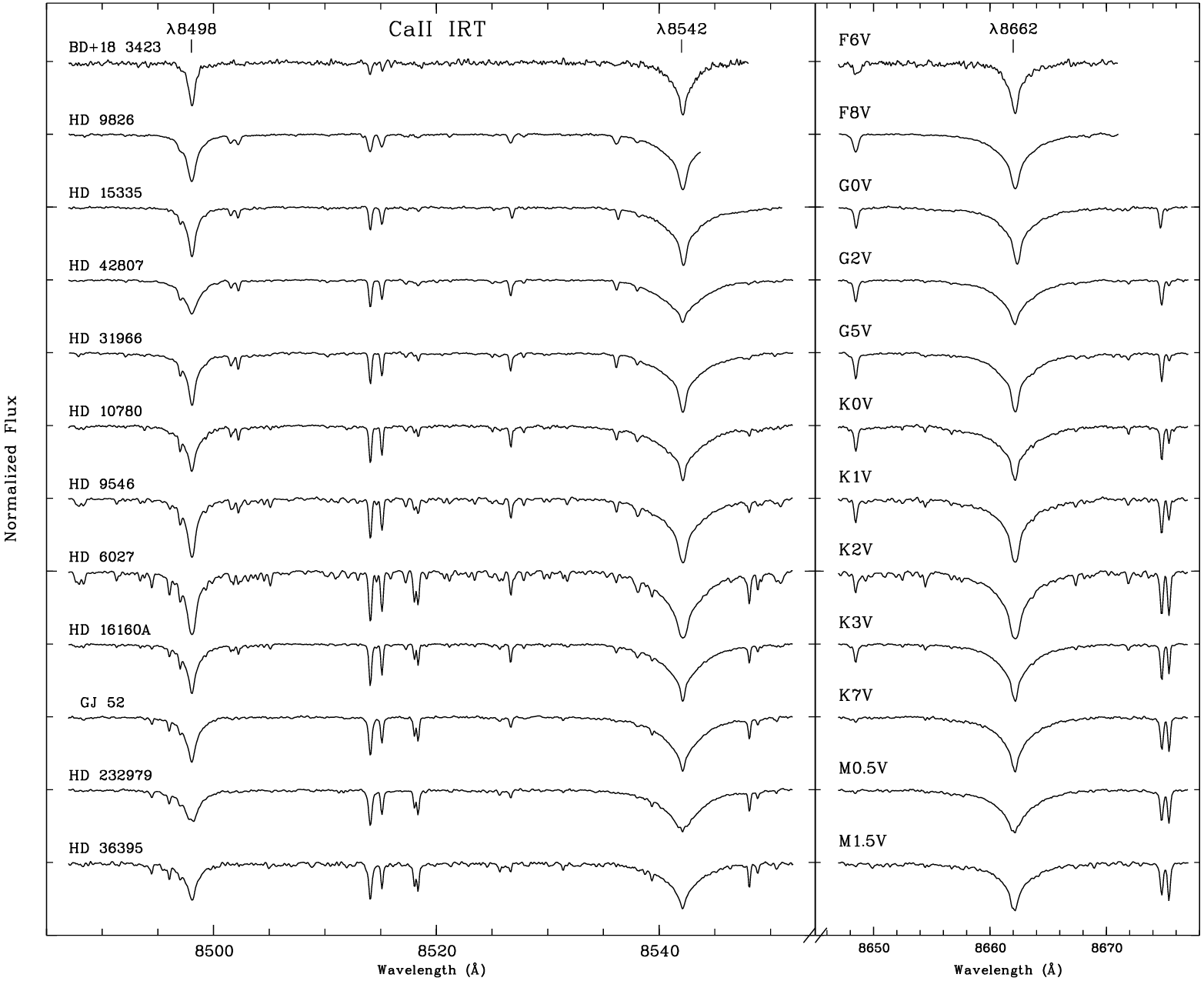,bbllx=34pt,bblly=164pt,bburx=665pt,bbury=678pt,height=22.4cm,width=17.8cm,clip=}}
\caption[ ]{Spectra in the Ca~{\sc ii} IRT 
$\lambda$8498, 8542, 8662 lines region
\label{fig:cairt}}
\end{figure*}
%%+++++++++++++++++++

%**********************************************************
\end{document}